\pdfoutput=1
\documentclass[doublecol]{epl2}
\usepackage[latin9]{inputenc}
\usepackage{graphicx}
\usepackage{esint}
\usepackage{subscript}
\usepackage[unicode=true,
 bookmarks=true,bookmarksnumbered=false,bookmarksopen=false,
 breaklinks=true,pdfborder={0 0 1},backref=false,colorlinks=false]
 {hyperref}
\hypersetup{pdftitle={Screened exchange dynamical mean field theory },
 pdfauthor={A van Roekeghem and S Biermann}}

\makeatletter

\newcommand{\lyxdot}{.}

\makeatother

\begin{document}

\title{Screened exchange dynamical mean field theory and its relation to
density functional theory: SrVO\textsubscript{3} and SrTiO\textsubscript{3}}

\shorttitle{Screened exchange DMFT and its relation to DFT: SrVO\textsubscript{3} and SrTiO\textsubscript{3}}

\author{Ambroise van Roekeghem\inst{1,2}\thanks{E-mail: \email{vanroeke@cpht.polytechnique.fr}}
and Silke Biermann\inst{2,3,4,5}\thanks{E-mail: \email{biermann@cpht.polytechnique.fr}}}

\shortauthor{A. van Roekeghem and S. Biermann}

\institute{
\inst{1} Beijing National Laboratory for Condensed Matter Physics, and Institute of Physics, Chinese Academy of Sciences, Beijing 100190, China\\
\inst{2} Centre de Physique Th\'eorique, Ecole Polytechnique, CNRS UMR 7644, 91128 Palaiseau, France\\
\inst{3} Coll\`ege de France, 11 place Marcelin Berthelot, 75005 Paris, France\\
\inst{4} European Theoretical Synchrotron Facility, Europe\\
\inst{5} Kavli Institute for Theoretical Physics, University of California, Santa Barbara, California 93106, USA}


\abstract{We present the first application of a recently proposed electronic structure scheme to transition metal oxides: screened exchange dynamical mean field theory includes non-local exchange beyond the local density approximation and dynamical correlations beyond standard dynamical mean field theory. Our results for the spectral function of SrVO$_{3}$ are in agreement with available experimental data, including photoemission spectroscopy and thermodynamics. Finally, the 3$d^{0}$ compound SrTiO$_{3}$ serves as a test case to illustrate how the theory reduces to the band structure of standard electronic structure techniques for weakly correlated compounds.}

\maketitle

\section{Introduction}

Combined density functional many body theories, such as the combination
of the local density approximation with dynamical mean field theory
(``LDA+DMFT'') \cite{LDA+DMFT-licht,LDA+DMFT-anisimov-1997,biermann_ldadmft},
have revolutionized the field of first principles electronic structure
calculations over the last decade, and many successful applications
to materials ranging from simple transition metals \cite{biermann_akw,licht_katsnelson_kotliar}
or f-electron metals \cite{Pu-kotliar-savrasov,PhysRevLett.87.276404,amadon:205112}
to complex oxides \cite{biermann:026404,Sr2IrO4-cyril} have been
worked out. Nevertheless, it has also been pointed out that conceptual
deficiencies related to the use of the Kohn-Sham band structure or
the neglect of dynamical screening effects \cite{udyn-michele,udyneff-michele,udyn-werner}
may have practical implications limiting the predictive power of the
approach. A more satisfactory approach, formulated entirely in a Green's
function language, is the combined many-body perturbation dynamical
mean field theory scheme ``GW+DMFT'' \cite{GW+DMFT-biermann,biermann-proc-2004a,biermann-proc-2004b}.
The first implementations of GW+DMFT have recently been elaborated,
with applications to SrVO\textsubscript{3} \cite{SrVO3-GW+DMFT-Jan,SrVO3-GW+DMFT-Jan-long}
and to systems of adatoms on surfaces \cite{GW+DMFT-surfaces-Philipp}.
Nevertheless, since the approach is computationally heavy, an important
line of research remains the developing and testing of approximate
schemes, and several different simplified schemes \cite{SrVO3-GW+DMFT-Held,SrVO3-GW+DMFT-Ferdi,Ambroise-BaCo2As2}
have recently appeared.

Here, we focus on the recently proposed ``Screened Exchange Dynamical
Mean Field Theory'' \cite{Ambroise-BaCo2As2}, where a one-particle
Hamiltonian beyond DFT-LDA is constructed and supplemented by a many-body
self-energy calculated from DMFT with frequency-dependent local Hubbard
interactions. We apply this new method to the transition metal oxide
SrVO\textsubscript{3}, a moderately correlated metallic compound
which -- thanks to its simple electronic structure -- has become a
benchmark compound for DMFT-based methods. The results are in agreement
with the available experimental information stemming from angle-resolved
and angle-integrated photoemission spectroscopy, optics and thermodynamics.
We confirm a striking band-widening effect in the unoccupied part
of the electronic structure, recently discovered within GW+DMFT \cite{SrVO3-GW+DMFT-Jan-long},
and which urgently calls for investigations by inverse photoemission
or related techniques. Finally, we apply our scheme to the weakly
correlated band insulator SrTiO\textsubscript{3}: here, a most interesting
cancellation of band widening by screened exchange and dynamical effects
elucidates the relation of the new technique to standard Density Functional Theory (DFT)
or GW techniques. Using (lightly doped) SrTiO\textsubscript{3} as a prototype
we argue that Screened Exchange Dynamical Mean Field Theory reproduces
the band structure of the standard theories for materials with weak
correlations.

\section{Screened exchange dynamical mean field theory}

Combining screened exchange with dynamical mean field theory with
frequency-dependent interactions (``dynamical DMFT'' or ``DDMFT'')
brings in two important corrections, as compared to standard LDA+DMFT,
namely (1) non-local exchange beyond the LDA and (2) dynamical screening
beyond usual DMFT. The scheme can be understood as a simplified version
of combined ``GW+DMFT'' or as a dynamical non-perturbative generalization
of Hedin's COulomb-Hole-Screened-EXchange (COHSEX) approximation \cite{Hedin-review}.
From a functional point of view, it corresponds to replacing the screened
Coulomb interaction in the nonlocal part of the GW+DMFT functional
(see e.g. Eq.~(4) in Ref.~\cite{GW+DMFT-biermann}) by its zero-frequency
value. In practice, in the recently implemented combined ``Screened
Exchange plus Dynamical DMFT'' (SEx+DDMFT) scheme \cite{Ambroise-BaCo2As2},
a one-particle Hamiltonian is constructed as 
\begin{equation}
H_{0}=H_{Hartree}+H_{SEx}=H_{LDA}-V_{xc}^{LDA}+H_{SEx}
\end{equation}
 where $H_{SEx}$ is a screened Fock exchange term, calculated from
the Yukawa potential 
\begin{equation}
\frac{e^{2}exp(-k_{TF}|r-r^{\prime}|)}{|r-r^{\prime}|}
\end{equation}
 with screening wavevector $k_{TF}$. The latter is self-consistently
calculated from the spectral function at the Fermi level. The density
used in the construction of $H_{0}$ is taken to be the converged
DFT-LDA one. The underlying rationale of this choice is the observation
that DFT-LDA densities are often extremely close to the true densities,
even when the Kohn-Sham band structure is not a good approximation
to spectral excitations.

The many-body part of the Hamiltonian 
\begin{equation}
H_{int}=H_{int\ stat}+H_{screening}
\end{equation}
 contains two terms: (1) static two-body interactions 
\begin{equation}
H_{int\ stat}=\frac{1}{2}\sum_{i}\sum_{m\sigma\neq m'\sigma'}V_{m\sigma m'\sigma'}n_{im\sigma}n_{im'\sigma'}
\end{equation}

as in usual LDA+DMFT, but with the \textit{bare} Coulomb interaction
matrix $V$ incorporating Hubbard interactions and Hund's coupling.
The sums $m,m^{\prime}$ run over the correlated orbitals, that is,
in our case the $t_{2g}$ states of SrVO\textsubscript{3}, localized
at atomic sites labeled by $i$.\\
 (2) Dynamical screening as described by introducing a set of auxiliary
screening bosons (creation and annihilation operators $b^{\dagger}$
and $b$), coupling to the physical electrons:

\begin{eqnarray}
H_{screening} & = & \sum_{i}\int_{\omega}d\omega\left[\lambda_{\omega}\sum_{m\sigma}n_{im\sigma}(b_{i,\omega}^{\dagger}+b_{i,\omega})\right.\nonumber \\
 & + & \left.\omega\left(b_{i,\omega}^{\dagger}b_{i,\omega}+\frac{1}{2}\right)\right].
\end{eqnarray}

The set of screening bosons can be understood as a parametrization
of local dynamical Coulomb interactions $\mathcal{U}(\omega)$, as
recently introduced in dynamical DMFT calculations \cite{udyneff-michele}.
Physically, these bosonic screening modes describe plasmons or particle-hole
excitations, which reduce the initial Coulomb interaction at low-energies,
making the effective interaction a frequency-dependent (retarded)
quantity. The final Hamiltonian $H=H_{0}+H_{int}$ has the form of
a multi-orbital Hubbard-Holstein Hamiltonian, which we treat within
dynamical mean field theory, using a continuous-time hybridization
expansion Monte Carlo solver \cite{TRIQS-website}. In this work,
we use the recent implementation of Ref.~\cite{Ambroise-BaCo2As2}.

In our applications to SrVO\textsubscript{3} and SrTiO\textsubscript{3}
below, we focus on the $t_{2g}$-states only, thus dealing with a
three-fold degenerate dynamical impurity problem. Calculations are
performed on the imaginary time axis at a temperature T = 400 K, and
analytically continued to obtain spectral functions by means of a
Maximum Entropy scheme \cite{MaxEnt}.

Due to the degenerate nature of the $t_{2g}$-states an important
simplification occurs for the self-consistent calculation of the screening
wavevector: indeed, Luttinger's theorem prevents any modifications of
the spectral function on the Fermi level by a local self-energy (other
than temperature-induced changes which are tiny in our case). Therefore,
the self-consistent value of the screening wavevector can already be obtained
from the result of the SEx calculation, resulting in $k_{TF}=1.1$
$a_{0}^{-1}$.

\section{SrVO\textsubscript{3}, a correlated metal}

SrVO\textsubscript{3} has become a classical ``drosophila'' compound
in the field of electronic structure calculations for correlated materials.
The reason is easy to see: the perfect cubic crystal structure of
this \textit{d}$^{1}$ compound results in a crystal field that lifts
the five-fold degeneracy of the V-3\textit{d} states, leaving only
the $t_{2g}$-states at the Fermi level. The low-energy electronic
structure is thus of appealing simplicity with a single electron in
a manifold of three perfectly degenerate $t_{2g}$-states. Transport
measurements have early on pointed out a surprisingly large Fermi
liquid regime, extending practically up to room temperature \cite{Onoda-resistivity}.
Due to the relatively localized nature of the corresponding orbitals
and a non-negligible Coulomb interaction, the spectral properties
are however not well-described by simple band theory. Various experimental
probes, in particular photoemission \cite{SrVO3-sekiyama,Mott-Hubbard-fujimori,Inoue-thesis,Maiti-CaSrVO3,Maiti-PRB-CaSrVO3,SrVO3-eguchi},
inverse photoemission \cite{SrVO3-morikawa} and angle-resolved photoemission
\cite{SrVO3-yoshida,Takizawa-SrVO3,Yoshimatsu-SrVO3,SrVO3-Yoshida-2010}
have been used to characterize the spectral properties of this compound.
The resulting picture is that of a moderately correlated metal, with
quasi-particle mass enhancement of the order of 2 and a well-defined
lower Hubbard band that has been identified in various spectroscopic
probes.

Electronic structure calculations have been able to reproduce this
correlated metal behavior, and the study of the weight transfer between
quasi-particle states and the lower Hubbard band has been one of the
early achievements of LDA+DMFT \cite{SrVO3-sekiyama,perovskite-pavarini,Ishida-SrVO3}.
Later on, more refined details of the electronic structure such as
possible kink structures in the low-energy electronic spectra \cite{Nekrasov-SrVO3}
or non-local correlations \cite{Valenti-SrVO3} have also come into
focus.

Despite the seemingly well-understood physics of SrVO\textsubscript{3},
the recent application of combined GW+DMFT reserved however new surprises:
it was found, for example, that a prominent peak observed in inverse
photoemission and usually interpreted as an upper Hubbard band of
$t_{2g}$-character should be dominantly attributed to the $e_{g}$-states
\cite{SrVO3-GW+DMFT-Jan-long}. The true upper Hubbard band is in
fact located at much lower energies, such that it merges with the
quasi-particle band structure observed in inverse photoemission.

A rather detailed overview of the experimental and theoretical investigations
has recently been given in Ref.~\cite{SrVO3-GW+DMFT-Jan-long}.

\begin{figure}
\begin{centering}
\includegraphics[width=8.5cm]{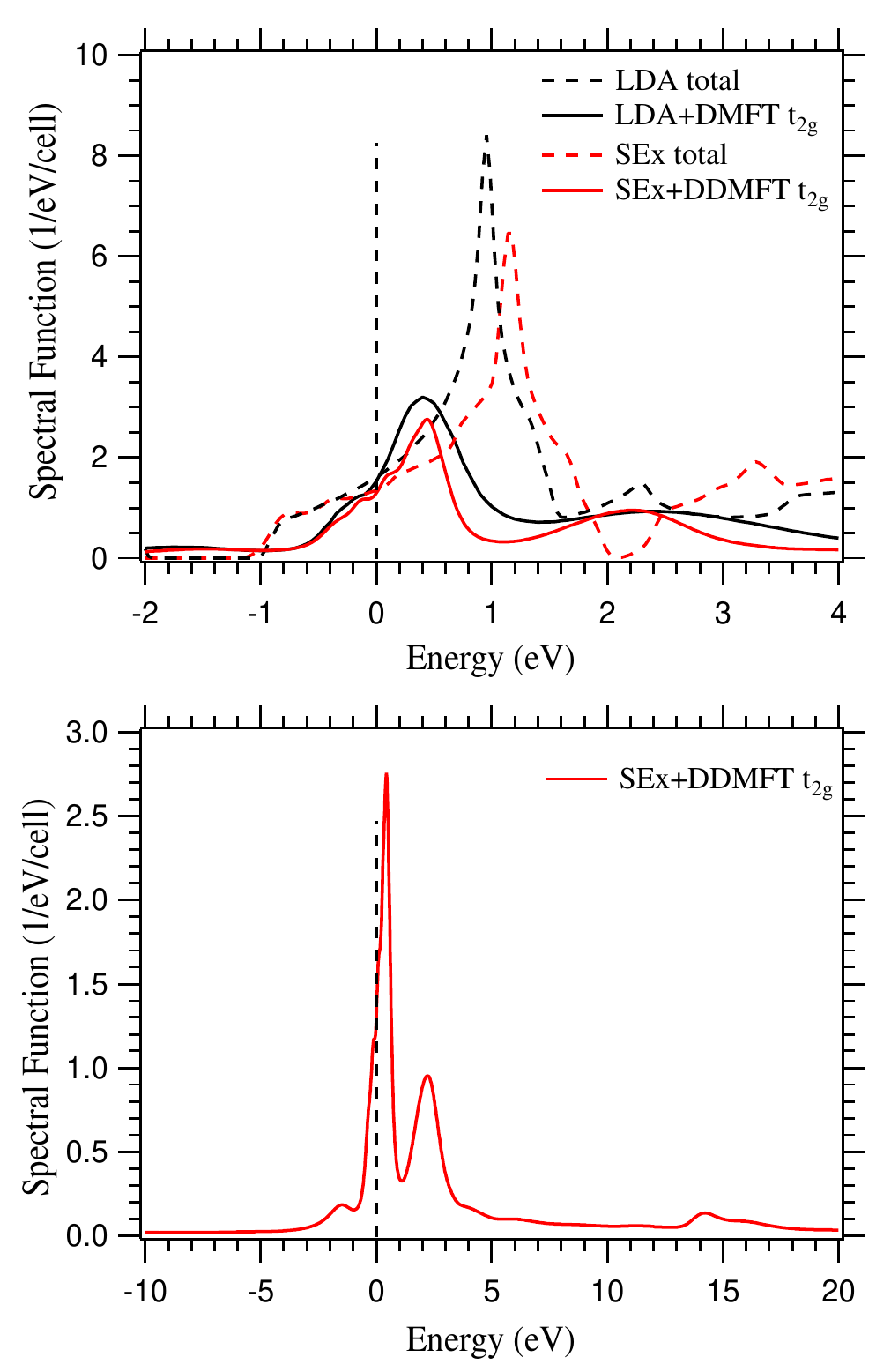} 
\par\end{centering}

\caption{Momentum-integrated spectral function of the $t_{2g}$ states of SrVO\textsubscript{3}
within SEx+DDMFT. Also given on the upper panel is the spectral function
of the $t_{2g}$ states from standard LDA+DMFT with U=4.0, reproduced
from \cite{Wannier-lechermann-2006}, the LDA total density of states
and the SEx total density of states. The larger energy scale in the
lower panel allows us to identify also the peaks due to the plasmon
excitation at 15 eV.\label{DOS1}}
\end{figure}

We now turn to the discussion of our results for SrVO\textsubscript{3}
within SEx+DDMFT. The momentum-integrated spectral function (Fig.~\ref{DOS1})
displays a three-peak structure -- a renormalized quasi-particle peak
and Hubbard bands -- as in LDA+DMFT. Indeed, in the occupied part
of the spectrum the picture is very close to the LDA+DMFT one. For
positive energies, however, significant differences are observed:
the upper Hubbard band is located at about 2 eV, whereas LDA+DMFT
calculations find it at about 2.5 eV. This difference corresponds
to what has also recently been found within GW+DMFT, which displays
a spectral function very similar to SEx+DDMFT. In these calculations,
it was revealed that a pronounced peak in inverse photoemission at
2.7 eV, which in the DMFT literature had been interpreted as the upper
Hubbard band of $t_{2g}$ character, is in fact dominantly made up
from $e_{g}$-states. It was speculated there that the only reason why
inverse photoemission has not identified the true upper Hubbard band
at 2 eV yet, is its close energetic position to a quite broad quasi-particle
peak, together with the very limited resolution of inverse photoemission
spectroscopy. Here, we confirm this picture within the new SEx+DDMFT
scheme. It will be most interesting to probe the unoccupied part of
the spectrum within inverse photoemission spectroscopy in order to
locate the 2 eV peak consistently found in GW+DMFT and SEx+DDMFT.

The lower panel of Fig.~\ref{DOS1} displays the same spectral function,
albeit on a larger energy scale. Here, additional features become
visible, namely plasmon replicas at 15 eV and, less pronounced, at
around 5 eV. These peaks stem from corresponding features in the frequency-dependent
effective Hubbard interactions, encoding the presence of collective
plasma oscillations \cite{SrVO3-GW+DMFT-Jan-long}.

\begin{figure}
\begin{centering}
\includegraphics[width=8.5cm]{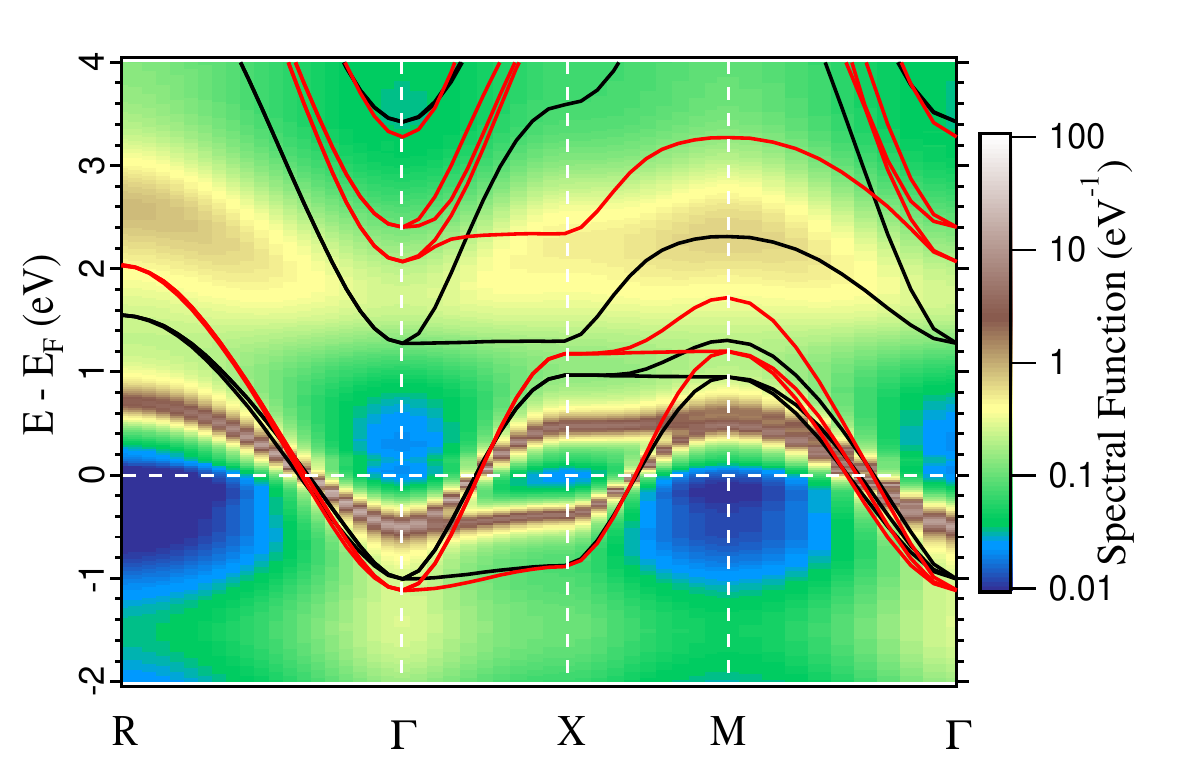} 
\par\end{centering}

\caption{Spectral function of the $t_{2g}$ states of SrVO\textsubscript{3}
within screened exchange dynamical mean field theory. Superimposed
are the LDA band structure (black lines), and the screened exchange
band structure (red lines).\label{Akw}}
\end{figure}

The momentum-resolved spectral function within screened exchange dynamical
mean field theory is shown in Fig.~\ref{Akw}, along with the LDA
band structure and the band structure obtained from the screened exchange
Hamiltonian. The central feature is the dispersing quasi-particle
band which has undergone a band renormalization of roughly a factor
of 2. Indeed, the bottom of the band at the $\Gamma$ point is found
at about -0.5 eV binding energy, in agreement with angle-resolved
photoemission data \cite{SrVO3-yoshida}. We determine the quasi-particle
weight corresponding to the DMFT self-energy as $Z^{-1}=1-\frac{\partial\Sigma_{DMFT}(i\omega)}{\partial i\omega}$
and find a value of 0.5. This reduction of spectral weight goes hand
in hand with a transfer to higher energies, and the lower {[}upper{]}
Hubbard bands are indeed clearly distinguished in particular at those
k-points, where the quasi-particle bands are occupied {[}empty{]}.
As expected from the k-integrated spectra, the upper Hubbard band
is located at around 2 eV, whereas the lower one is found between
-1 and -2 eV.

\begin{figure}
\begin{centering}
\includegraphics[width=8.5cm]{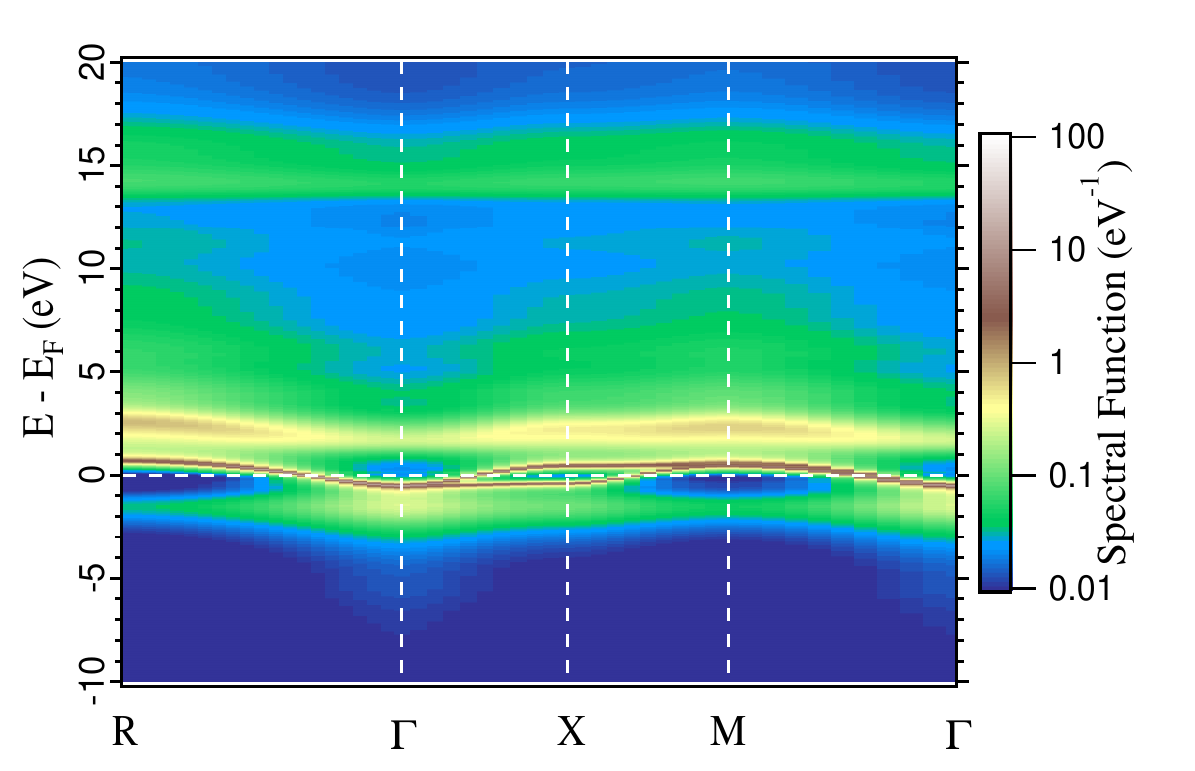} 
\par\end{centering}

\caption{Spectral function of the $t_{2g}$ states of SrVO\textsubscript{3}
within screened exchange dynamical mean field theory, as in Fig.\ \ref{Akw},
but on a larger energy scale.\label{Akw2}}
\end{figure}

Also plotted are the LDA Kohn-Sham band structure, and the eigenvalues
of the screened exchange Hamiltonian $H_{0}$. Here, interesting effects
of the screened exchange are identified. As usual, exchange tends
to widen the band \cite{Ashcroft-Mermin,SrVO3_nonlocal_cyril}, however,
as recently also found in GW+DMFT spectra, this effect is larger in
the unoccupied part of the spectrum. For this reason, GW+DMFT and
SEx+DDMFT spectra are close to LDA+DMFT ones in their occupied parts,
whereas larger corrections are obtained for empty states. In Fig.~\ref{Akw2},
we plot the momentum-resolved spectral function on a larger energy
scale, where also the high-energy plasmon at 15 eV can be clearly
distinguished.

\textbf{Electron-doped SrTiO3} Finally, we discuss a most interesting
implication of our theory for weakly correlated materials. Indeed,
both screened exchange and the dynamical tail of an otherwise irrelevant
Hubbard interaction will still have an effect on the spectral function
of a weakly correlated material, the former as a band broadener and
the latter as a band narrower, as discussed above. On the example
of very lightly electron-doped SrTiO\textsubscript{3}, we demonstrate
that these effects largely cancel and result in a band structure extremely
close to the DFT Kohn-Sham one. SrTiO\textsubscript{3} is a well-studied
band insulator where the only true deficiency of the Kohn-Sham band
structure is believed to be the too small band gap opening between
oxygen \textit{p}-states and Ti-3\textit{d} empty bands. The shape
of the latter, however, is largely reproduced even if more advanced
theories are applied: the GW approximation offers a band gap correction,
without notably modifying the $t_{2g}$ band structure itself \cite{Amadon-GW-SrTiO3,Vanderbilt-GW-SrTiO3}.
Here, we will be interested neither in the gap nor in the rich physics
resulting from electron-lattice coupling in SrTiO\textsubscript{3},
but rather focus on the band structure of the $t_{2g}$ bands, for
which we will show that SEx+DDMFT reproduces to a good approximation
the shape and width of the Kohn-Sham bands of DFT. To this effect,
we will consider the slightly electron-doped compound, where the $t{}_{2g}$-bands
are occupied with 0.05 electrons per Ti, and we use the experimental
cubic lattice parameter a = 3.905 \AA{}.

\begin{figure}
\begin{centering}
\includegraphics[width=8.8cm]{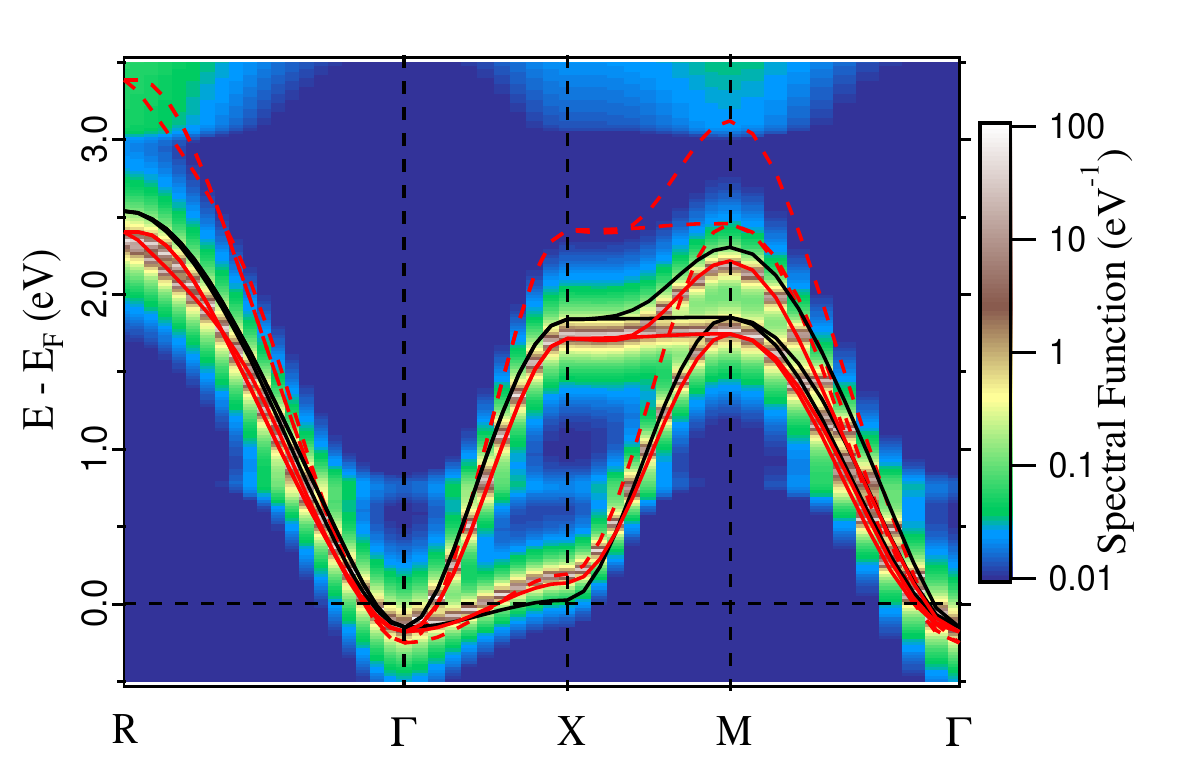} 
\par\end{centering}

\caption{Band structure of the $t_{2g}$ states of SrTiO\textsubscript{3}
within LDA (black lines), Screened Exchange (red dashes) and SEx renormalized
by a plasmonic factor $Z_{B}=0.7$ (red lines) superimposed on the
SEx+DDMFT $t_{2g}$ spectral function. The chemical potential corresponds
to $n=0.05$ electron doping per Ti atom, which gives a self-consistent
Thomas-Fermi screening wavevector of $k_{TF}=0.6$ $a_{0}^{-1}$ according
to the SEx density of states.\label{SrTiO3}}
\end{figure}

In Fig.~\ref{SrTiO3}, we compare the LDA band structure for the
$t_{2g}$ states to the SEx one and to the band structure for the
$t_{2g}$ states resulting from dynamical narrowing effects of the
SEx bands. In Ref.~\cite{udyneff-michele}, it was shown that the
effect of the dynamic high-energy tail of the interaction can -- in
the antiadiabatic limit -- be understood in terms of a global band
renormalization factor that can directly be calculated from the frequency-dependence
of the screened Hubbard interaction: $Z_{B}=\exp(-\int_{0}^{+\infty}d\omega\Im U(\omega)/(\pi\omega^{2}))$.
In SrVO\textsubscript{3}, the value of the bosonic renormalization
factor calculated using the interaction from constrained-RPA is $Z_{B}=0.7$
\cite{udyneff-michele}. The value we calculate for electron-doped
SrTiO\textsubscript{3} (with 0.05 electrons) based on the undoped
LDA bandstructure on which we apply a chemical potential shift is
the same within the numerical precision. While the SEx bandwidth is
1.36 times larger than LDA, after we renormalize the SEx bands by
$Z_{B}=0.7$ the bandwidth is comparable to LDA. This band structure
is in agreement with the experimental inverse photoemission spectrum,
which displays a peak around 1.7--2.0 eV \cite{Tezuka-BIS-SrTiO3,Sarma-BIS-SrTiO3,Shin-BIS-SrTiO3,Baba-BIS-SrTiO3}
corresponding to the flat band along the XM direction. The band structure
of lightly electron-doped SrTiO\textsubscript{3} has also been measured
by angle-resolved photoemission in Ref.~\cite{Rotenberg-ARPES-SrTiO3,Santander-Nature-ARPES-SrTiO3}
and the estimates of the effective masses along the $\Gamma$X direction
-- 0.7$m_{e}$-1.2$m_{e}$ for the light electrons and 7.0$m_{e}$-20$m_{e}$
for the heavy electrons -- are consistent with the LDA band dispersion
at the bottom of the bands, further confirming the above picture.

These results indicate that at this very low doping level electron-electron
scattering due to the instantaneous part of the Hubbard interaction
is negligible, and the width and shape of the band is essentially
determined by the counteracting effects of screened exchange and dynamical
screening. We now turn to the full SEx+DDMFT calculation for this
lightly electron-doped SrTiO\textsubscript{3}, see colored background
of Fig.\ \ref{SrTiO3}, which indeed confirms this scenario: the
maxima of the spectral function correspond to the SEx bands renormalized
by $Z_{B}$ (``SEx+$Z_{B}$''). The case of electron-doped SrTiO\textsubscript{3}
thus strongly suggests that the remarkable performance of LDA in the weakly
correlated regime is largely due to to error cancellations between
non-local exchange and dynamical screening effects.

However, while the LDA bandstructure seems to portray relatively well
the quasiparticles dispersions, the underlying physics is somewhat
simplistic: the reduction of the low-energy spectral weight which
necessarily accompanies the band renormalization by $Z_{B}$ is beyond
the LDA, and the description of the high-energy plasmonic excitations
-- seen for instance in SrTiO\textsubscript{3} \cite{Tezuka-BIS-SrTiO3,Higuchi-SXES-SrTiO3}
-- requires a full SEx+DDMFT calculation.

\section{Conclusion and perspectives}

We have performed combined screened exchange dynamical mean field
calculations for the ternary transition metal oxides SrVO\textsubscript{3}
and electron-doped SrTiO\textsubscript{3}. The results for SrVO\textsubscript{3}
are consistent with full GW+DMFT calculations \cite{SrVO3-GW+DMFT-Jan-long}.
In particular, we confirm here a pronounced band widening effect on
the unoccupied part of the spectrum by the screened exchange part
of the Hamiltonian. The final spectral function after the DMFT calculation
resembles the one of standard LDA+DMFT calculations in the occupied
part of the spectrum, but substantial modifications are obtained in
the conduction band. It will be most interesting to probe empty states
of SrVO\textsubscript{3} -- and more generally of correlated materials
-- within momentum-resolved spectroscopic probes to confirm these
findings.

In lightly electron-doped SrTiO\textsubscript{3}, we have shown how
the theory reduces to a band picture given that the dynamical tail
can be integrated in the formation of electronic polarons and that
the probability of double occupancy at this low filling level is negligible.
The antagonistic effects of the non-local screened exchange and of
the high-frequency tail of the Hubbard interaction are such that the
quasiparticles dispersions are close to the DFT-LDA result. Still,
the underlying physics of plasmonic excitations is ignored in LDA,
in contrast to our SEx+DDMFT scheme. Our findings shed light on the
apparent success not only of combined LDA+DMFT techniques, but also
of the Kohn-Sham band structure itself, when used as an approximation
to excitation energies for materials with weak electronic Coulomb
correlations.

\acknowledgments

We acknowledge useful discussions with R. Claessen, I. Mazin,
A. Santander, and the authors of Ref.~\cite{Ambroise-BaCo2As2}
and \cite{SrVO3-GW+DMFT-Jan-long}, in particular Thomas Ayral, Michel
Ferrero and Olivier Parcollet for support and development of the TRIQS
toolkit \cite{TRIQS-website}. This work was supported by IDRIS/GENCI
Orsay under project 091393, the National Science Foundation under
Grant No. NSF PHY11-25915, and the European Research Council under
project 617196.

\bibliographystyle{eplbib}
\bibliography{BiblioFirstprinciples,BiblioU,Biblio113,Biblio214,BiblioDFT,BiblioWannier,BiblioGeneral,BiblioDMFT,BiblioGW,Bibliopnictides,BiblioARPES,refs}

\end{document}